\documentclass[prb,reprint,
superscriptaddress,nofootinbib, amsmath,amssymb,aps,english,showpacs,longbibliography,floatfix,]{revtex4-2} 

\usepackage{graphicx} 

\usepackage[dvipsnames]{xcolor}
\usepackage{amsmath,amsfonts}
\usepackage{comment}
\usepackage{ifthen}
\usepackage{enumerate}
\usepackage{enumitem}

\newboolean{redactswitch}
\setboolean{redactswitch}{false} 

\newcommand{\redact}[1]{\ifthenelse{
    \boolean{redactswitch}}{{[Redacted]}}{
    {#1}}}

\newcommand{\fracroot}[2]{\ifthenelse{#1=1}{\frac{1}{\sqrt{#2}}}{\sqrt{\frac{#1}{#2}}}}

\newif\ifshowtimestamp
\showtimestampfalse

\usepackage{xcolor}
\usepackage{etoolbox}
\usepackage{soulutf8} 
\usepackage[normalem]{ulem}
\sethlcolor{yellow}
\soulregister\cite7
\soulregister\ref7
\soulregister\textbf7

\newtoggle{review}
\togglefalse{review} 

\newcommand{\add}[1]{%
  \iftoggle{review}
    {\hl{#1}}
    {#1}%
}

\newcommand{\del}[1]{%
  \iftoggle{review}
    {\textcolor{red}{\sout{#1}}}
    {}%
}

\newcommand{\newentry}[1]{
    \iftoggle{review}{\textcolor{violet}{#1}}{#1}
}

\usepackage[colorlinks=true,urlcolor=blue,citecolor=blue,linkcolor=blue]{hyperref}		

\hypersetup{colorlinks=true,urlcolor=blue,citecolor=blue,linkcolor=blue}

\date{\today}

\begin{document}

\title{Resource Letter QIE-1: Research in quantum information education}

\author{Josephine C. Meyer}
\email{jmeyer26@gmu.edu}
\affiliation{Dept.\ of Physics and Astronomy, George Mason University, Fairfax, VA 22030, USA}

\author{Simon Goorney}
\affiliation{Dept.\ of Management, Aarhus University, Aarhus, Denmark}

\author{Tunde Kushimo}
\affiliation{Dept.\ of Physics, Wichita State University, Wichita, KS 67208, USA}

\author{Zeki C. Seskir}
\affiliation{Institute for Technology Assessment and Systems Analysis, Karlsruhe Institute of Technology, Karlsruhe, Germany}

\begin{abstract}
    In celebration of the UN International Year of Quantum Science and Technology, this Resource Letter surveys the rapidly-growing field of scholarship in quantum information science and engineering (QISE) education. It is primarily written as a guide for educators wishing to get started teaching QISE using research-based teaching methods, as well as for discipline-based education research (DBER) practitioners looking to get started in this field. Topics covered include scoping the field of QISE education, research into student reasoning in QISE, research-based and research-inspired curricular materials from the high school to graduate level, research-based assessments, simulation and gamification tools, and tools for incorporating discussion of the societal and ethical implications of quantum technologies into the classroom.
\end{abstract}

\maketitle

\section{Introduction}

The year 2025 \del{is}\add{was} declared the UN International Year of Quantum Science and Technology \cite{UN:2024}, and educators throughout the world are seizing the opportunity to incorporate principles from quantum information science and engineering (QISE)\footnote{The field is variously called Quantum Information Science (QIS), Quantum Engineering, Quantum Technologies, Quantum Information Science and Technology (QIST), and increasingly even -- much to the consternation of some physicists -- just ``Quantum.'' By adopting the QISE acronym, we are choosing a common convention but without taking a stance on the field's official name.} into their curricula. Until recently, QISE topics were largely taught only in graduate-level physics electives; now, there is a push to teach them at the high school level and even before. In response, a robust and growing body of literature from the physics and engineering education communities on how to teach QISE topics effectively has begun to coalesce, though at present resources tend to be scattered across a variety of journals (in both physics and engineering education) presenting a particular challenge for educators or new discipline-based education research (DBER) practitioners wishing to become familiar with this literature.

Most of the articles cited in this Resource Letter were written in the past 3-5 years. The field of QISE education research is still rapidly growing and evolving, and new and important literature appears on the arXiv pre-print server seemingly every week. \del{Accordingly, t}\add{T}he purpose of this Letter is not to provide a definitive forecast of what QISE education research will become, but rather to simply provide a snapshot of the field as it stands at \del{time of writing (fall 2025)}\add{the close of the International Year of Quantum.} \textbf{\add{Accordingly, all resources included within this resource letter have a cutoff date of December 31, 2025.}}\footnote{\add{Note:  articles disseminated as preprints prior to the December 31, 2025 cutoff were considered to have met this threshold, regardless of formal publication status. In all such cases, the version of record was cited if available at the time this Resource Letter went to press, occasionally resulting in a 2026-dated citation.}}

\add{We emphasize that as with any Resource Letter, this list is intended to be neither exhaustive or definitive. Its scope is influenced both by editorial decisions (e.g.~the choice to prioritize formal education research methodologies) and by structural gaps in the literature (see Sec.~\ref{sec:missing}}).

As authors, we believe the field of QISE education has seemingly endless directions to continue exploring, and we wish this Resource Letter above all to serve as an inspiration for researchers and educators to become involved in shaping the field while it remains new and in flux. As such, we conclude with a brief call for future research, noting several possible avenues for work in this field. Nothing would bring us more delight than if this Letter spearheads enough new research that a follow-up Resource Letter QIE-2 becomes warranted within this decade.

\subsection{What is quantum information science and engineering (QISE)?}

QISE is a new and emerging field at the intersection of physics, math, electrical engineering, computer science, and beyond seeking to harness the principles of quantum mechanics to circumvent classical limitations on information processing. Whereas Quantum 1.0 technologies (e.g.\ lasers, semiconductor chips) leveraged quantum mechanics to improve the performance and capabilities of classical information processing, Quantum 2.0 technologies encode information itself in quantum form. As of \del{2025}\add{2026}, the main quantum information technologies are typically considered to be quantum computing (including quantum simulation), quantum communications, and quantum sensing. The theory and implementations of quantum information technologies are well-covered in Resource Letter QI-1: Quantum Information \cite{Strauch:2016}.


\subsection{Methodology and criteria for article inclusion}
\label{sec:inclusion}

This project was born out of informal discussions within the quantum education and Responsible Quantum Technologies \cite{Schmidt:2025} research communities about the need for a curated list of readings accessible to educators and beginning researchers. All four co-authors on this project are early-career researchers (graduate students or postdocs) who have become known for pioneering work on QISE education. We presently hail from 4 different institutions across the US and Europe, and were born and raised in 4 different countries across 3 continents. 
Given our unique disciplinary backgrounds, geographic locations, and social identities, we believe each of us carries a unique perspective to contribute to this Resource Letter. 

Accordingly, we adopted a consensus-based approach for the inclusion of articles in this Resource Letter. As a starting point, we compiled resources from the bibliography \redact{of first author \redact{JCM's} thesis (published in spring 2025 \cite{Meyer:2025})}, chosen as a starting point because it represented, to our knowledge, one of the most comprehensive QISE education resource lists available to date. Additional articles and resources not in \redact{the original thesis bibliography} were then nominated by the authors to create an initial pool of 167 articles.

A voting form was sent to each author\add{ in September 2025}, and each author ranked each article on a 5-point Likert scale (ranging from -2, ``strong reject,'' to +2, ``strong include'') abstaining from any articles that they were co-authors on or otherwise had a conflict of interest reviewing. As a first pass, articles with an average score of +1 or higher (or multiple ``strong include'') and no ``strong rejects'' were selected for inclusion in the Resource Letter; articles with a score below 0.5 (or multiple ``strong rejects'') and no ``strong include'' were eliminated. The remaining articles were then discussed as a group to obtain consensus, with an eye toward ensuring the voices of any who had voted ``strong reject'' or ``strong accept'' on an article were adequately heard. The final resource list presented here is a product of this consensus voting process. A small number of additional articles were added subsequently by informal consensus that either were nominated too late to incorporate into the voting form\del{ or}\add{, }were discovered through bibliometric analysis (Sec.~\ref{sec:bibliometrics})\add{, were published after the original September 2025 voting date, or were recommended by reviewers}.

\add{Note that our scope generally excludes resources related to teaching and learning of traditional quantum mechanics (QM), except where the resources were deemed of especially high relevance to teaching QISE topics. Teaching and learning of QM is its own long-established research tradition in Physics Education Research, and likely warrants its own Resource Letter.}

\subsection{Bibliometric analysis}
\label{sec:bibliometrics}

\begin{figure}
    \centering
    \includegraphics[width=1\linewidth]{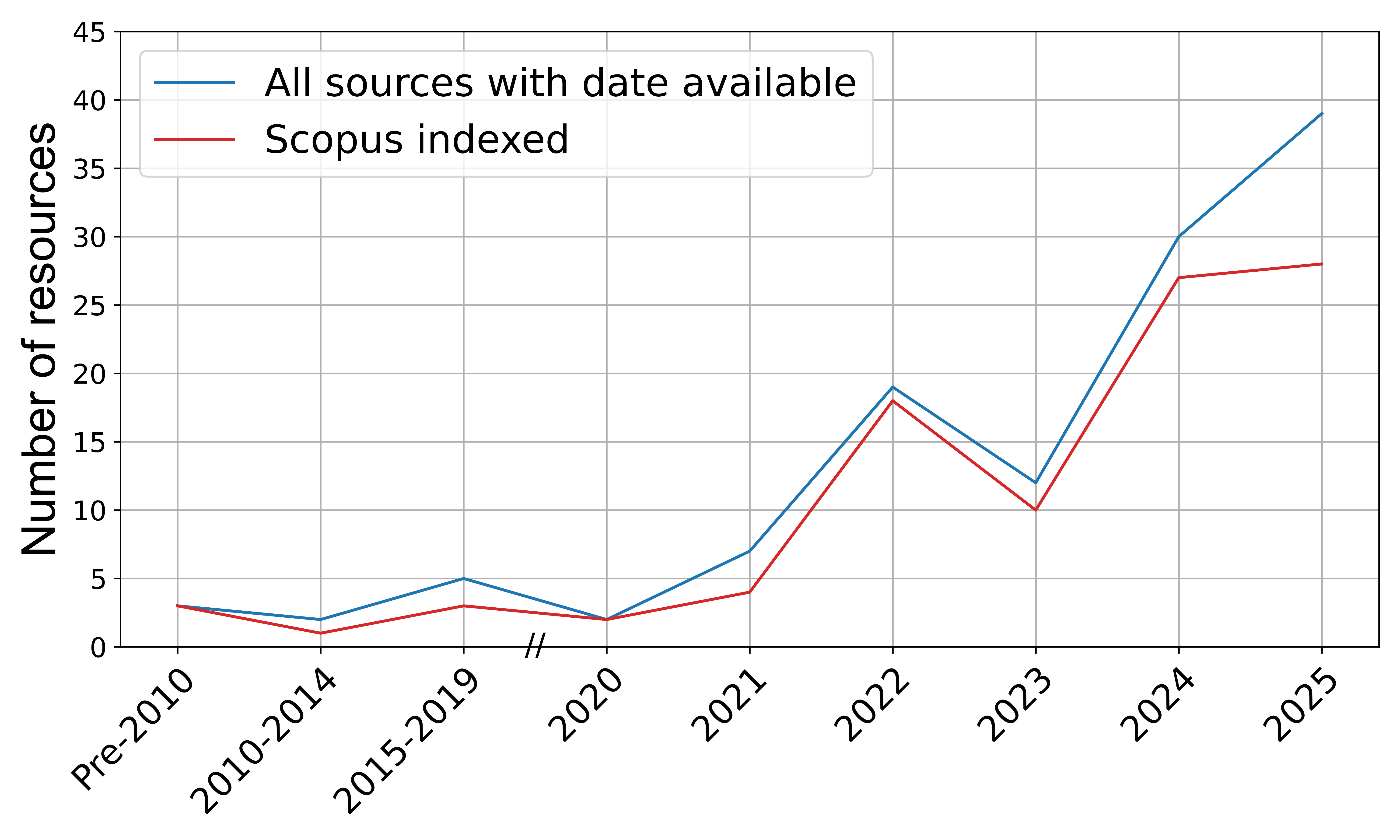}
    \caption{Comparison of total entries in our dataset vs.\ Scopus indexed resources by year. Note the abrupt change in $x$ axis scale and the inflection point in approximately 2020-2021. \add{Five resources with a 2026 publication date (typically released as preprints or early access in 2025), are not shown.}}
    \label{fig:1}
\end{figure}

For completeness sake, we performed bibliometric analysis of our dataset to identify whether we were missing any high impact resources. \add{Prior to initial submission, in October 2025, w}\del{W}e manually created a list of our resources on Scopus, downloaded its metadata and performed citation analysis using VOSviewer software \cite{vanEck2010} to identify sources highly cited by publications in our dataset. In total we identified 29 publications that were cited 4 or more times in the publications of our dataset. Out of these 29, 16 were already in our original dataset and we investigated the rest manually, after which we decided to include four and identified the remaining nine as not suitable for this resource letter due to their focus being entirely on QM and not QISE. We also compared the publication years of the total number of resources in our dataset against the Scopus indexed subset of our dataset, which resulted in Figure~\ref{fig:1} \add{(updated 2026)}.

The gap in 2024 and more so in 2025 is due to the high number of pre-print (mostly arXiv) publications in our resources. This, combined with the steep rise in number of publications starting from 2020 onwards, hints us that the field of QISE education is rapidly growing and becoming distinct from quantum mechanics education.

\subsection{What's missing from the literature?}
\label{sec:missing}

In constructing this Resource Letter, we have aimed to be as thorough as we can be while accurately reflecting the state of the literature. However, we note that some important structural gaps exist in the literature we were unable to address through curation:

\begin{itemize}
    \item Quantum sensing education: The papers presented here are overwhelmingly focused on quantum computing, with a secondary focus on quantum communications and networks. However, quantum sensing -- arguably the most mature quantum information technology at time of writing \cite{Gschwendtner:2024} -- is seldom\del{ if ever} discussed in the quantum education literature. 
    \item Diversity of student backgrounds: The vast majority of studies in the dataset were conducted in the US or Europe at elite institutions. If not addressed, this trend risks entrenching a ``quantum divide'' in access to quantum literacy and quantum jobs as the benefits of education research flow primarily to relatively privileged students \cite{TenHolter:2022,Gercek:2025}. 
    More research is needed to determine the extent to which education research findings translate to students in the Global South and/or at less-resourced universities.
    \item Despite the interdisciplinary nature of QISE coursework \cite{Meyer:2022}, the vast majority of research studies we found have been published in physics journals or conference proceedings.\footnote{The one important exception being the proceedings of the annual Quantum Science and Engineering Education Conference (QSEEC), a subconference of the IEEE International Conference on Quantum Computing and Engineering.} This disparity is especially apparent for high-quality DBER studies and might owe in part to journal editorial policies.
\end{itemize}

\subsection{Resource accessibility rankings}
Per standard practice for AJP Resource Letters, all articles and other resources are classified by their accessibility as follows:

\begin{itemize}
    \item \textbf{Elementary (E):} Accessible, instructor- or policymaker-focused articles with little-to-no background in QISE or education research required. An excellent starting point for those new to the field.
    \item \textbf{Intermediate (I):} Useful articles that may require familiarity with QISE content and/or education research theories and methodologies to understand in full. Those new to the field may wish to stick to the discussion and conclusions.
    \item \textbf{Advanced (A):} These resources will primarily be of value to education researchers.
\end{itemize}
\add{In general, within each section, articles are sorted by accessibility class (E, I, A), then by year and finally alphabetically by author. Liberty is taken where necessary and appropriate (e.g.~to group related studies together).}

\textbf{\add{Unless explicitly specified, no resources (regardless of classification) require familiarity with QISE content beyond the level typically encountered in a middle-division undergraduate QISE course.}}

\section{Resources for framing QISE education}

\subsection{QISE education landscape and educator studies}

A number of articles focus on understanding the QISE education landscape as it currently exists. Some of these studies interview instructors and other stakeholders while others are based on analysis of publicly-available datasets such as course catalogs.

\begin{enumerate}

    \item ``United States quantum education landscape,'' B.M.\ Zwickl \textit{et al.} \url{https://quantumlandscape.streamlit.app/}. An intuitive and up-to-date visualization of QISE (and traditional quantum mechanics) coursework and degree programs at US universities, showcasing the distribution and breadth of these programs. (E)

    \item ``Defining the quantum workforce landscape: A review of global quantum education initiatives,'' M.\ Kaur and A.\ Venegas-Gomez. \href{https://doi.org/10.1117/1.OE.61.8.081806}{Opt.\ Eng.} \textbf{61}(8), 081806 (2022). Notable for the study's focus on academic-industry alignment and for including nontraditional education pathways such as hackathons, conferences, and MOOCs. (E)

    \item ``Quantum technology master's: A shortcut to the quantum industry?'' S.\ Goorney \textit{et al.} \href{https://doi.org/10.1140/epjqt/s40507-024-00299-x}{EPJ Quantum Technol.} \textbf{12}, 2 (2025). A global landscape study of QISE master's programs, showing a marked shift over time from focus on academic to industry-focused careers and from physics to interdisciplinary faculties. (E)

    \item ``Landscape of quantum information science and engineering education: From physics foundations to interdisciplinary frontiers,'' A.R.\ Piña \textit{et al.} \href{https://doi.org/10.1103/pstt-46b9}{\del{arXiv:2504.13719}\add{Phys.~Rev.~Phys.~Educ.~Res.} \add{\textbf{21}, 020131} (2025)}. A detailed, accessible analysis of the distribution of QISE courses and degree programs in the US, with attention to equity and opportunities to incorporate QISE within existing coursework. (E)

    \item ``Today's interdisciplinary quantum information classroom: Themes from a survey of quantum information science instructors,'' J.C.\ Meyer \textit{et al.} \href{https://doi.org/10.1103/PhysRevPhysEducRes.18.010150}{Phys.\ Rev.\ Phys.\ Educ.\ Res.} \textbf{18}, 010150 (2022). A mixed-methods study of interdisciplinary QISE educators' experiences, focusing on the diversity of student and instructor backgrounds, prerequisites, curricular materials, and needs/opportunities especially useful for incoming education researchers. (I) 

    \item ``Building Europe's quantum technology education community,'' S.\ Goorney, E.\ Karydi, and J.\ Sherson. \href{https://doi.org/10.1140/epjqt/s40507-025-00362-1}{EPJ Quantum Technol.} \textbf{12}, 61 (2025). A detailed discussion of the European quantum education community, focused on an activity-theoretic analysis of 11 pilot projects. (I)

    \item ``Preparing students for the quantum information revolution: Interdisciplinary teaching, curriculum development, and advising in quantum information science and engineering,'' F.\ \del{Seiholli}\add{Seifollahi} and C.\ Singh. \href{https://doi.org/10.1088/1361-6404/ae0200}{Eur.\ J.\ Phys.} (2025). A qualitative study featuring 7 interviews with university-level QISE instructors of diverse disciplinary backgrounds. The interviewees reflect on the challenges and opportunities of teaching and mentoring across disciplines. (I)

    \item \newentry{``Building bridges in quantum information science education: Expert insights to guide framework development for interdisciplinary teaching and evolution of common language,'' L.\ Doyle, F.\ Seifollahi, and C.\ Singh. \href{https://doi.org/10.1140/epjqt/s40507-025-00454-y}{EPJ Quantum Technol.} \textbf{13}, 2 (2026). Shares reflections from university-level QISE instructors on the ongoing evolution of a common interdisciplinary language for describing quantum concepts across disciplines, as well as appropriate levels of abstraction for introducing core QISE concepts in interdisciplinary introductory coursework. (I)}

    \item ``Disparities in access to US quantum information education,'' J.C.\ Meyer, G.\ Passante, and B.R.\ Wilcox. \href{https://doi.org/10.1103/PhysRevPhysEducRes.20.010131}{Phys.\ Rev.\ Phys.\ Educ.\ Res.} \textbf{20}, 010131 (2024). A quantitative study of QISE course availability at US institutions, demonstrating that coursework is heavily concentrated at elite, well-resourced institutions in heavily urbanized states and calling for initiatives to improve access at low-income and rural schools. (A) 

    \begin{itemize}
    
    \item The authors published an accessible summary of the study's findings in \href{https://theconversation.com/unequal-access-to-quantum-information-education-may-limit-progress-in-this-emerging-field-now-is-the-time-to-improve-231565}{The Conversation}. (E)
    \end{itemize}

\end{enumerate}

\subsection{QISE education policy}

\add{Resources focusing on policies and opportunities for growth within the QISE education landscape, of particular interest to policymakers and program administrators:}

\begin{enumerate}[resume]

    \item ``Quantum Information Science and Technology Workforce Development National Strategic Plan,'' \href{https://www.quantum.gov/wp-content/uploads/2022/02/QIST-Natl-Workforce-Plan.pdf}{U.S.\ National Science \& Technology Council} (2022). The plan outlines a coordinated federal strategy to assess QIST workforce needs, expand access through outreach and education, and address quantum-training gaps in quantum-relevant careers. (E)

    \item \newentry{``Quantum education in Africa: Innovative models, collaborative frameworks, and the path forward,'' F.\ Mazhandu and M.\ Mafu. \href{https://doi.org/10.1109/QCE65121.2025.20523}{Proc.\ 2025 IEEE Int.\ Conf.\ Quantum Comput.\ Eng.} \textbf{3}, 65-69 (2025). An accessible analysis of the state of QISE education and research across the African continent, along with policy recommendations. (E)}

    \item ``UK Quantum Skills Taskforce Report,'' \href{https://www.gov.uk/government/publications/uk-quantum-skills-taskforce-report/uk-quantum-skills-taskforce-report}{UK Department for Science, Innovation \& Technology} (2025). Sets out current and future skills demands in the UK quantum sector and proposes 16 recommendations to align education, training pathways, technical roles, and equity goals to build a sustainable quantum workforce. (I)

    \item ``Insights from educators on building a more cohesive quantum information science and engineering education ecosystem,'' S.\ El-Adawy \textit{et al.} \del{arXiv:2507.01578}\href{https://doi.org/10.1103/tfmb-hnvz}{\add{Phys.~Rev.~Phys.~Educ.~Res.}} \add{\textbf{21}, 020144} (2025). The authors analyze interviews with educators at 15 institutions to identify common challenges and strategies in developing QISE programs, using a SWOT framework to highlight issues of alignment, collaboration, evaluation, and stakeholder connections in building a cohesive quantum education ecosystem. (I)

    \item ``Investigating opportunities for growth and increased diversity in quantum information science and engineering education in the U.S.,'' A.R.\ Piña \textit{et al.} \del{arXiv:2505.00104 }\href{https://doi.org/10.18260/1-2--56888}{\add{Proc.~2025 ASEE Ann.~Conf.~Expo.}}\add{, 48011} (2025). This work analyzes over 8,000 course listings and nearly 90 dedicated QISE programs in the U.S., revealing concentration in research-intensive institutions and advocating for broader integration of quantum content into engineering curricula to expand access and diversity. (I)

    \item ``Outcomes from a workshop on a national center for quantum education,'' E.\ Barnes \textit{et al.} \href{https://doi.org/10.1140/epjqt/s40507-025-00343-4}{EPJ Quantum Technol.} \textbf{12}, 40 (2025). It synthesizes outcomes from a multi-stakeholder NSF-sponsored workshop on establishing a national center for quantum education, identifying coordinated strategies for educator training, curriculum development, infrastructure, and evaluation to strengthen quantum pathways and broaden participation. (A)

\end{enumerate}

\subsection{Quantum workforce needs}

\add{These studies explore the needs of companies in the quantum industry, useful for educators and policymakers seeking academia-industry alignment.}

\begin{enumerate}[resume]

    \item ``Preparing for the quantum revolution: What is the role of higher education?'' M.F.J.\ Fox, B.M.\ Zwickl, and H.J.\ Lewandowski. \href{https://doi.org/10.1103/PhysRevPhysEducRes.16.020131}{Phys.\ Rev.\ Phys.\ Educ.\ Res.} \textbf{16}, 020131 (2020). \del{Fox et al. (2020) is} An early study in which 21 U.S. companies were interviewed regarding the skills they value in different job roles. Skills such as coding, data analysis, and laboratory experience, were found to be rather universal across roles. As of the time of writing, PhD graduates were still dominant in the job market, but later research articles have shown this is beginning to change. (E)

    \item ``Achieving a quantum smart workforce,'' C.\ Aiello \textit{et al.} \href{https://doi.org/10.1088/2058-9565/abfa64}{Quantum Sci.\ Technol.} \textbf{6}(3), 030501 (2021). Is a 2021 perspective article reporting on the status of 18 early efforts in university education for QISE.The authors advocate for integrating quantum concepts into existing STEM programs, providing quantum knowledge for broader audiences as QISE becomes more commercialised. (E)

    \item ``Assessing the needs of the quantum industry,'' C.\ Hughes \textit{et al.} \href{https://doi.org/10.1109/TE.2022.3153841}{IEEE Trans.\ Educ.} \textbf{65}(4), 592-601 (2022). \del{Hughes et al. (2022) investigated} \add{Investigates} what kind of jobs, skills, and degrees are needed for the development of the quntum industry, through a survey of 57 companies. They identified that many job roles are highly technical and require a PhD-level education, but there are also a substantial number of roles, even rather technical ones, which offer positions for bachelor and master graduates. The authors comment that many STEM degrees should offer some quantum courses, making these job roles accessible to a wide number of graduates. (E)

    \item ``Industry perspectives on projected quantum workforce needs,'' S.\ El-Adawy \textit{et al.} \del{arXiv:2508.15055}\href{https://doi.org/10.1119/perc.2025.pr.El-Adawy}{\add{Proc.~2025 Phys.~Educ.~Res.~Conf.}}\add{, 148-153} (2025). \del{"This is} An interview study investigating how quantum industry professionals in managerial positions feel about workforce needs. The authors found that managers anticipate a need for a range of educational levels from bachelors to PhDs in physics, engineering, and computer science to fill the roles available in industry. Managers also believe that there will be an increased need for individuals who can apply quantum information science knowledge across fields. (E)
\del{"}

    \item ``Industry insights into quantum knowledge needed for the quantum information science and engineering workforce,'' A.R.\ Piña \textit{et al.} \href{https://doi.org/10.1119/perc.2025.pr.Pina}{\del{arXiv:2509.15039}\add{Proc.~2025 Phys.~Educ.~Res.~Conf.}}\add{, 350-355} (2025). This interview study was undertaken with managers and employees at 21 QISE companies. The authors identified job roles associated with different quantum requirements - awareness, conversational, proficient, and expert. The findings are presented as preliminary, but they provide some insight into the level of expertise required for different kinds of quantum jobs in 2025. (E)

    \item \newentry{``Quantum Workforce Report Series,'' S.\ El-Adawy, A.R.\ Piña, H.J.\ Lewandowski, and B.M.\ Zwickl (2025-2026). A set of 3 whitepapers on the roles, skills, and job duties of a broad cross-section of the US quantum industry, designed to be easily accessible to nonexperts. (E)}

    \begin{itemize}
        \item \textbf{Report \#1:} ``Experimental skills for non-Ph.D.\ roles in the quantum industry'' \href{https://arxiv.org/abs/2510.12936}{arXiv:2510.12936}
        \item \textbf{Report \#2:} ``Categorization of roles in the quantum industry'' \href{https://arxiv.org/abs/2511.11820}{arXiv:2511.11820}
        \item \textbf{Report \#3:} ``Profiles of roles in the quantum industry''\footnote{This third report narrowly missed the 31 Dec.\ 2025 cutoff, but is included for completeness of the series.} \href{https://arxiv.org/abs/2601.04292}{arXiv:2601.04292}
    \end{itemize}

    \item ``The quantum technology job market: Data driven analysis of 3641 job posts,'' S.\ Goorney \textit{et al.} \del{arXiv:2503.19004}\href{https://doi.org/10.1140/epjqt/s40507-026-00477-z}{\add{EPJ Quantum Technol.}} (202\del{5}\add{6}). This is the first large-scale empirical study of the quantum job market, using data from 3,641 online job postings. The authors identify trends across geography, sectors, company sizes, and degree requirements, and highlight tensions around brain drain and brain gain between countries. (E)

    \item ``Quantum technician skills and competencies for the emerging Quantum 2.0 industry,'' M.\ Hasanovic \textit{et al.} \href{https://doi.org/10.1117/1.OE.61.8.081803}{Opt.\ Eng.} \textbf{61}(8), 081803 (2022). This paper argues that technician education is a missing but essential link in the quantum workforce pipeline, and identify 5 key areas for technican education. (I)

    \item ``Mapping quantum industry demands to education: A critical analysis of skills, qualifications, and modalities,'' S.\ Devendrabrabu, S.\ Ganguly, and K.\ Hemachandran. \href{https://doi.org/10.1140/epjqt/s40507-025-00406-6}{EPJ Quantum Technol.} \textbf{12}, 105 (2025). This article investigates job descriptions from 90 quantum jobs in detail, identifying skills, degrees, and qualifications mentioned in the job descriptions. They compare among different quantum applications and hardware platforms, identifying that each hardware modality requires a strong basis in quantum physics and its own specialized set of skills. (\del{E/}I)

\end{enumerate}

\del{A particularly valuable set of resources are associated with t} The European Competence Framework for Quantum Technologies\del{,} \add{is} a toolkit developed in the EU to catalog QISE-related skills and match educational programs to skills and industry needs. While developed specifically in the European context, the findings are broadly applicable across international contexts. The full link to the latest version of the Competence Framework (which undergoes frequent modification and refinement) can be found here: \url{https://zenodo.org/records/15209861}.

Additionally, a number of articles have been published by the team responsible for developing and maintaining the Competence Framework. Several \add{of the }articles discuss aspects of the Framework's development. \del{In addition, they also do}\add{All }include valuable findings related to the quantum workforce, whether one is interested in adopting the Competence Framework or not:

\begin{enumerate}[resume]
    \item ``Requirements for future quantum workforce -- a Delphi study,'' F.\ Gerke\footnote{Author now publishes under the name Franziska Greinert.} \textit{et al.} \href{https://doi.org/10.1088/1742-6596/2297/1/012017}{J.\ Phys.:\ Conf.\ Ser.} \textbf{2297} (GIREP Malta Webinar 2020), 012017 (2022). (E)

    \item ``Future quantum workforce: Competences, requirements, and forecasts,'' F.\ Greinert \textit{et al.} \href{https://doi.org/10.1103/PhysRevPhysEducRes.19.010137}{Phys.\ Rev.\ Phys.\ Educ.\ Res.} \textbf{19}, 010137 (2023). (E)

    \item ``Towards a quantum ready workforce: The updated European Competence Framework for Quantum Technologies,'' F.\ Greinert \textit{et al.} \href{https://doi.org/10.3389/frqst.2023.1225733}{Front.\ Quantum Sci.\ Technol.} \textbf{2}, 1225733 (2023). (E)

    \item ``Advancing quantum technology workforce: Industry insights into qualification and training needs,'' F.\ Greinert \textit{et al.} \href{https://doi.org/10.1140/epjqt/s40507-024-00294-2}{EPJ Quantum Technol.} \textbf{11}, 82 (2024). (E)

    \item ``Extending the European Competence Framework for Quantum Technologies: New proficiency triangle and qualification profiles,'' F.\ Greinert \textit{et al.} \href{https://doi.org/10.1140/epjqt/s40507-024-00302-5}{EPJ Quantum Technol.} \textbf{12}, 1 (2025). (E)

\end{enumerate}

\add{Note that the European Competence Framework has been successfully applied outside the original European context, for instance:}

\begin{enumerate}[resume]
    \item \newentry{``Quantum computer engineering education in the Philippines: A readiness and recommendation analysis of the prescribed computer engineering curriculum,'' D.J.\ Domingo Lopez \textit{et al.} \href{https://doi.org/10.1109/QCE65121.2025.20525}{Proc.\ 2025 IEEE Int.\ Conf.\ Quantum Comput.\ Eng.} \textbf{3}, 71-80 (2025). Analyzes the existing undergraduate computer engineering curriculum in the Philippines to identify areas of alignment and recommendations for improvement to ensure that graduates of the program are prepared for future quantum workforce jobs (I).}
\end{enumerate}

\subsection{Diversity in quantum}

\add{High-profile educators and researchers (e.g.~\cite{Aiello:2021a}) have argued that the Second Quantum Revolution presents a once-in-a-generation opportunity to address longstanding diversity and culture issues in STEM. These articles provide additional context and actionable suggestions:}

\begin{enumerate}[resume]

    \item ``The quantum computer revolution must include women,'' C.\ Singh. \href{https://www.scientificamerican.com/article/the-quantum-computer-revolution-must-include-women/}{Scientific American} (2021 Jan 13). \del{Singh (2021) shows}\add{Argues } that women report a weaker sense of belonging and self efficacy in the field, and suggests a number of strategies to help improve feelings of integration for women in QISE. (E)

    \item ``Q-Turn: Changing paradigms in quantum science,'' A.B.\ Sainz. \href{https://doi.org/10.1088/2058-9565/ac82c4}{Quantum Sci.\ Technol.} \textbf{7}(4), 044004 (2022). Sainz (2022) presents Q-Turn, an international workshop series that integrates quantum information science with social awareness and community reform. The workshops show\del{s} potential for higher participation by women, non-binary scientists, and researchers from the Global South than typical in quantum conferences. (E)

    \item ``Creating a modular, workforce-relevant undergraduate curriculum for quantum information science and engineering for all people,'' C.D.\ Porter, Z.\ Atiq, and E.\ Fletecher. \href{https://doi.org/10.1119/perc.2022.pr.Porter}{Proc.\ 2022 Phys.\ Educ.\ Res.\ Conf.}, 365-370 (2022). Describes the QuSTEAM initiative, a large-scale collaborative effort among research universities, community colleges, and HBCUs to design an inclusive, modular undergraduate curriculum in QISE. (E)
    
    \item ``Reflections of quantum educators on strategies to diversify the Second Quantum Revolution,'' A.\ Ghimire and C.\ Singh. \href{https://doi.org/10.1119/5.0223983}{Phys.\ Teach.} \textbf{63}, 35-39 (2025). This article explores how QISE might avoid inheriting the inequitable culture of traditional physics. Educators emphasize introducing quantum ideas early in K–12 and undergraduate curricula, highlighting diverse applications, and explicitly teaching confidence and belonging alongside competence. (E)
    
    \item ``Leveraging equity and inclusion to demystify and diversify undergraduate quantum science education,'' M.\ Mafu and F.\ Mazhandu. \href{https://doi.org/10.1119/5.0231016}{Phys.\ Teach.} \textbf{63}, 90-94 (2025). Discusses the interconnected structures (social, economic, and power) which can affect diversity in and access to QISE. (E)

\end{enumerate}

\subsection{Useful case reports}

These case studies focus on individual educational initiatives but have important conclusions for the field as a whole.

\begin{enumerate}[resume]

    \item ``Quantum computing online workshops and hackathon for Spanish speakers: A case study,'' A.\ Maldonado-Romo and L.\ Yeh. \href{https://doi.org/10.1109/QCE53715.2022.00096}{Proc.\ 2022 IEEE Int.\ Conf.\ Quantum Comput.\ Eng.}, 709-717 (2022). Reports on a Spanish-language online workshop series and hackathon in Latin America, highlighting strong demand for accessible quantum computing resources, participant learning outcomes, and the value of culturally and linguistically inclusive educational initiatives. (E)

    \item \newentry{``Leveraging dual enrollment programs to expand secondary education in quantum computation,'' D.L.\ Tucker. \href{https://doi.org/10.1109/QCE57702.2023.20319}{Proc.\ 2023 IEEE Int.\ Conf.\ Quantum Comput.\ Eng.} \textbf{3}, 10-14 (2023). Illustrates how partnerships between universities and regional high schools can scalably expand access to quantum computing education at the high school level, with particular attention paid to reducing barriers to entry for underresourced schools. (E)}

    \item ``Introducing quantum information and computation to a broader audience with MOOCs at OpenHPI,'' G.\ Hellstern, J.\ Hettel, and B.\ Just. \href{https://doi.org/10.1140/epjqt/s40507-024-00270-w}{EPJ Quantum Technol.} \textbf{11}, 59 (2024). Analyzes a sequence of nine MOOCs on quantum computing offered via OpenHPI, reporting on participant demographics, learning behaviors, and success rates to show how modular, low-threshold online pathways can broaden access to quantum education. (E)

    \item ``Quantum Entrepeneurship Lab: Training a future workforce for the quantum industry,'' A.\ Sander \textit{et al.} \del{arXiv:2505.06298}\href{https://doi.org/10.1109/QCE65121.2025.20550}{\add{Proc.~2025 IEEE Int.~ Conf.~Quantum Comput.~Eng.}}\add{ \textbf{3}, 185-193} (2025). Presents the Quantum Entrepreneurship Lab, a one-semester, project-based course at Technical University of Munich (TUM) designed to bridge academia and industry by pairing technical and business students to turn quantum research ideas into potential commercial ventures. (E)

    \item \newentry{``Quantum workshop for IT professionals,'' B.\ Just, J.\ Hettel, and G.\ Hellstern. \href{https://doi.org/10.1140/epjqt/s40507-025-00450-2}{EPJ Quantum Technol.} \textbf{13}, 4 (2026). Presents the design and evaluation of a one-day module intended to upskill existing industry professionals, and demonstrates that an approach based primarily on use cases and a business-simulation activity rather than physical/mathematical formalism is successful in meeting the needs and expectations of these audiences. (E)}

    \item ``Building a quantum engineering undergraduate program,'' A.\ Asfaw \textit{et al.} \href{https://doi.org/10.1109/TE.2022.3144943}{IEEE Trans.\ Educ.} \textbf{65}(2), 220-242 (2022). A community-developed roadmap for undergraduate quantum engineering education, detailing course frameworks, degree tracks, and training opportunities to prepare both ``quantum-aware'' and ``quantum-proficient'' engineers. (I)

    \item ``Development of an undergraduate quantum engineering degree,'' A.S.\ Dzurak \textit{et al.} \href{https://doi.org/10.1109/TQE.2022.3157338}{IEEE Trans.\ Quantum Eng.} \textbf{3}, 6500110 (2022). Describes the creation and first year of operation of one of the world’s first undergraduate degree in quantum engineering, offering a model for integrating quantum mechanics with engineering curricula to address anticipated workforce demand. (I)

    \item ``The design and implementation of a quantum information science undergraduate program,'' S.\ Blanchette \textit{et al.} \href{https://arxiv.org/abs/2412.01874}{arXiv:2412.01874} (2024). Describes the design and launch of a 3.5-year undergraduate program in quantum information science at Université de Sherbrooke, detailing curriculum choices and institutional ecosystem influences to help bridge the quantum workforce gap. (I)

    \item ``The quantum technology open master: Widening access to the quantum industry,'' S.\ Goorney, M.\ Sarantinou, and J.\ Sherson. \href{https://doi.org/10.1140/epjqt/s40507-024-00217-1}{EPJ Quantum Technol.} \textbf{11}, 7 (2024). The authors conceptualize and evaluate an ``Open Master'' transnational education model in quantum technology, showing how local accreditation mechanisms and community-building strategies can widen access to specialist skills and help institutions expand teaching capacity. (I)

    \item ``QCaMP: A 4-week summer camp introducing high school students to quantum information science and technology,'' M.\ Ivory \textit{et al.} \del{arXiv:2504.15977}\href{https://doi.org/10.1109/QCE65121.2025.20510}{\add{Proc.~2025 IEEE Int.~Conf.~Quantum Comput.~Eng.}}\add{ \textbf{3}, 29-38} (2025). Presents the design of a project-based QISE summer camp for high school students, integrating hands-on modules, professional development, and exposure to quantum career pathways. (I)

    \item ``From computing to quantum mechanics: accessible and hands-on quantum computing education for high school students,'' Q.\ Sun \textit{et al.} \href{https://doi.org/10.1140/epjqt/s40507-024-00271-9}{EPJ Quantum Technol.} 
    \textbf{11}, 58 (2024). Narrates practical learning in QISE through the use of portable NMR quantum computers, demonstrating how hands-on experiences can enhance students’ engagement and conceptual understanding. (I)

    \item ``Project-based learning in introductory quantum computing courses: A case study on quantum algorithms for medical imaging,'' N.B.\ Gautam, K.E.\ Schubert, and E.P.\ Blair. \del{arXiv:2508.21321}\href{https://doi.org/10.1109/QCE65121.2025.20511}{\add{Proc.~2025 IEEE Int.~Conf.~Quantum Comput.~Eng.}}\add{ \textbf{3}, 39-50} (2025). Illustrates how a project-based learning module centered on quantum algorithms for medical imaging (e.g.\add{~}HHL vs.\add{~}ART) can help students move beyond passive understanding to engage with implementation, critique, and interdisciplinary connections. (I)

    \item \newentry{``Bridging the theory-practice gap: A university course on quantum information science and transfer to career-oriented physics teaching,'' M.\ Förster and G.\ Pospiech. \href{https://doi.org/10.1140/epjqt/s40507-025-00441-3}{EPJ Quantum Technol.} \textbf{12}, 127 (2025). A novel approach to teacher training, bridging an internship with traditional university coursework, demonstrates potential for helping prospective teachers align instruction with industry needs. (I)}
    
    \item ``The USRA Quantum Feynman Academy: If you give a student a quantum internship,'' Z.\ Gonzalez Izquierdo \textit{et al.} \href{https://arxiv.org/abs/2505.04641}{arXiv:2505.04641} (2025). The authors present an eight-year retrospective on the USRA Feynman Quantum Academy internship program, analyzing how sustained experiential training supports student development, community building, and pathways into quantum careers. (I)


\end{enumerate}

\section{Research on teaching and learning QISE concepts}

Numerous studies have begun to investigate the teaching and learning of QISE across a variety of disciplinary backgrounds. These studies use numerous methodologies and focus on a variety of diverse academic and disciplinary boundaries. The following two articles provide useful background framing on QISE education research and are recommended for all readers new to the field:

\begin{enumerate}[resume]
    \item ``Advancing quantum information science pre-college education: The case for learning sciences collaboration,'' R.\ Coelho \textit{et al.} \del{arXiv:2508.00668}\href{https://doi.org/10.1109/QCE65121.2025.20508}{\add{Proc.~2025 IEEE Int.~Conf.~Quantum Comput.~Eng.}}\add{ \textbf{3}, 11-22} (2025). A position paper arguing for the importance of cooperation between educators and the education research community to advance QISE education. Applicable to all levels, not just K-12. (E)

    \item ``A framework for curriculum transformation in quantum information science and technology education,'' S.\ Goorney \textit{et al.} \href{https://doi.org/10.1088/1361-6404/ad7e60}{Eur.\ J.\ Phys.} \textbf{45}(6), 065702 (2024). A practical toolkit for development of research-inspired curricula in QISE, distilling the essential findings of decades' worth of STEM education research into a framework easily applied by educators with concrete examples. (E)
\end{enumerate}

\add{A 2024 meta-analysis of research-based interventions in quantum teaching and learning is also available:}

\begin{enumerate}[resume]
    \item \newentry{``Empirical insights into the effects of research-based teaching strategies in quantum education.'' A.\ Donhauser \textit{et al.} \href{https://doi.org/10.1103/PhysRevPhysEducRes.20.020601}{Phys.\ Rev.\ Phys.\ Educ.\ Res.} \textbf{20}, 020601 (2024). A systematized review of the literature on research-based strategies for teaching and learning quantum concepts, both traditional QM and QIS, with an eye toward quantum technologies applications. Primarily of interest to education researchers. (I)}
\end{enumerate}

\subsection{Research on student conceptual learning and reasoning}

\add{A key goal of education research is to understand how students think and reason, e.g.~as a starting point for designing research-based curricular and assessment instruments. Several of these studies directly investigate student learning in QISE contexts; the remainder, though situated in (e.g.) general QM courses, were included for their close connection to central QISE topics:}

\begin{enumerate}[resume]

    \item ``Student ability to differentiate between superposition and mixed states,'' G.\ Passante, P.J.\ Emigh, and P.S.\ Shaffer. \href{https://doi.org/10.1103/PhysRevSTPER.11.020135}{Phys.\ Rev.\ ST Phys.\ Educ.\ Res.} \textbf{11}, 020135 (2015). Investigates how undergraduate and graduate students understand the distinction between superposition and mixed states, revealing some persistent difficulties that can inform QISE instructional design. (I)

    \item ``Investigating students' strategies for interpreting quantum states in an upper-division quantum computing course,'' J.C.\ Meyer \textit{et al.} \href{https://doi.org/10.1119/perc.2021.pr.Meyer}{Proc.\ 2021 Phys.\ Educ.\ Res.\ Conf.}, 289-294 (2021). Examines how upper-division QISE students interpret quantum states, particularly distinguishing superposition from mixed states. Two problem-solving strategies, Naive Measurement Probabilities (NMP) and Virtual Quantum Computer (VQC), were identified. (I)

    \item ``Investigating student interpretations of the differences between classical and quantum computers: Are quantum computers just analog classical computers?'' J.C.\ Meyer \textit{et al.} \href{https://doi.org/10.1119/perc.2022.pr.Meyer}{Proc.\ 2022 Phys.\ Educ.\ Res.\ Conf.}, 317-322 (2022). Investigates how upper-division and graduate-level QISE students reason about distinctions between classical and quantum computers, highlighting common misconceptions and conceptual challenges. (I)

    \item ``Investigating students' strengths and difficulties in quantum computing,'' T.\ Kushimo and B.\ Thacker. \href{https://doi.org/10.1109/QCE57702.2023.20322}{Proc.\ 2023 IEEE Int.\ Conf.\ Quantum Comput.\ Eng.} \textbf{3}, 33-39 (2023). Examines undergraduate students’ strengths and difficulties in an introductory quantum computing course to guide the development of research-based instructional materials. (I)

    \item ``Design and evaluation of a questionnaire to assess learners' understanding of quantum measurement in different two-state contexts: The context matters,'' P.\ Bitzenbauer \textit{et al.} \href{https://doi.org/10.1103/PhysRevPhysEducRes.20.020136}{Phys.\ Rev.\ Phys.\ Educ.\ Res.} \textbf{20}, 020136 (2024). Assesses learners’ understanding of quantum measurement across varied two-state contexts, offering a reliable tool for both classroom use and education research. (I)

    \item ``Keep it secret, keep it safe: Teaching quantum key distribution in high school,'' E.Y.\ Weissman \textit{et al.} \href{https://doi.org/10.1140/epjqt/s40507-024-00276-4}{EPJ Quantum Technol.} \textbf{11}, 64 (2024). Explores the feasibility of teaching quantum key distribution (QKD) at the high school level by testing varied teaching approaches. (I)

    \item ``Secondary and university students' descriptions of quantum superposition,'' N.B.C.\ Birkeland and M.V.\ Bøe. \href{https://doi.org/10.1119/5.0225893}{Phys.\ Teach.} \textbf{63}, 32-34 (2025). A comparative study of Norwegian upper secondary and university students’ descriptions of quantum superposition after engaging with digital learning resources. (I)

    \item ``Visualization enhances problem solving in multi-qubit systems: An eye-tracking study,'' J.\ Bley \textit{et al.} \del{arXiv:2505.21508}\href{https://doi.org/10.1103/hqzq-fbrb}{Phys.\ Rev.\ Phys.\ Educ.\ Res.} \textbf{22}, 010126 (2026). An investigation into how visualization supports learning in QISE, particularly for understanding Hadamard gate operations in multi-qubit systems. (I)

    \item ``From research to resources: Assessing student understanding and skills in quantum computing,'' B.\ Thacker \textit{et al.} \del{arXiv:2507/09656}\href{https://doi.org/10.1088/1361-6404/ae21a4}{\add{Eur.~J.~Phys.}} \add{\textbf{47}, 015702 }(2025). A comparative study examining students in introductory quantum computing courses taught with and without research-based mini-tutorials, offering valuable insights for both new and experienced educators. (I)

    \item ``Framework for understanding the patterns of student difficulties in quantum mechanics,'' E.\ Marshman and C.\ Singh. \href{https://doi.org/10.1103/PhysRevSTPER.11.020119}{Phys.\ Rev.\ ST Phys.\ Educ.\ Res.} \textbf{11}, 020119 (2015). Presents a framework for analyzing student reasoning difficulties and the development of expertise in quantum mechanics by drawing parallels with introductory classical mechanics.\add{ Several but not all of the discussed difficulties are pertinent to QISE specifically.} (A)

    \item ``Effect of an introductory quantum physics course using experiments with heralded photons on pre-university students' conceptions about quantum physics,'' P.\ Bitzenbauer. \href{https://doi.org/10.1103/PhysRevPhysEducRes.17.020103}{Phys.\ Rev.\ Phys.\ Educ.\ Res.} \textbf{17}, 020103 (2021).\add{ A quasi-controlled study demonstrating that explicit introduction of quantum optics in secondary school was more effective than the traditional German curricular framework in promoting a shift from mechanistic to probabilistic (quantumlike) reasoning about quantum mechanical quantities. (A)}

    \item ``Mathematical sense making of quantum phenomena using Dirac notation: Its effect on secondary school students' functional thinking about photons,'' F.\ Hennig \textit{et al.} \href{https://doi.org/10.1140/epjqt/s40507-024-00274-6}{EPJ Quantum Technol.} \textbf{11}, 61 (2024). Investigates whether introducing reduced Dirac notation in teaching quantum optics can foster secondary students’ functional thinking about photons. (A)

\end{enumerate}

Within this category are three studies that have specifically examined students' understanding of entanglement:

\begin{enumerate}[resume]

    \item ``Investigating student understanding of quantum entanglement,'' A.\ Kohnle and E.\ Deffebach. \href{https://doi.org/10.1119/perc.2015.pr.038}{Proc.\ 2015 Phys.\ Educ.\ Res.\ Conf.}, 171-174 (2015). This study uses the QuVis Entanglement simulation to examine undergraduate understanding of quantum entanglement, highlighting some persistent misconceptions. (I)

    \item ``Spooky action at a distance? A two-phase study into learners' views of quantum entanglement,'' M.\ Brang \textit{et al.} \href{https://doi.org/10.1140/epjqt/s40507-024-00244-y}{EPJ Quantum Technol.} \textbf{11}, 33 (2024). This is a two-phase empirical study examining pre-service physics teachers’ and university students’ conceptions of quantum entanglement, highlighting some persistent misconceptions. (I)

    \item ``Investigating students' understanding of entanglement,'' B.M.\ Zwickl and H.J.\ Hersom. \href{https://doi.org/10.1119/perc.2024.pr.Zwickl}{Proc.\ 2024 Phys.\ Educ.\ Res.\ Conf.}, 467-472 (2024). A study exploring how undergraduates conceptualize quantum entanglement, suggesting that reliance on classical analogies in textbooks can oversimplify the concept. (I)

\end{enumerate}

\subsection{Research on quantum labs}

Until recently, QISE-related laboratory courses have received little attention in the literature. Fortunately, this trend has begun to change in recent years with a flurry of papers by Borish and Lewandowski. Most commonly, these labs use single-photon quantum optics experiments.

\begin{enumerate}[resume]

    \item ``A hands-on introduction to single photons and quantum mechanics for undergraduates,'' B.\ Pearson and D.\ Jackson. \href{https://doi.org/10.1119/1.3354986}{Am.\ J.\ Phys.} \textbf{78}, 471-484 (2010). A useful primer on single-photon quantum optics experiments intended for instructors wishing to implement these experiments. (E)
    
    \item ``Implementation and goals of quantum optics experiments in undergraduate instructional labs,'' V.\ Borish and H.\ Lewandowski. \href{https://doi.org/10.1103/PhysRevPhysEducRes.19.010117}{Phys.\ Rev.\ Phys.\ Educ.\ Res.} \textbf{19}, 010117 (2023). An accessible introduction to the possible experiments, learning goals, and uses of single-photon experiments in upper-division undergraduate lab courses, based on a survey and interview study of instructors. Identifies creative adaptations used by educators to overcome challenges; particularly useful for instructors. (E)

    \item ``Seeing quantum effects in experiments,''  V.\ Borish and H.\ Lewandowski. \href{https://doi.org/10.1103/PhysRevPhysEducRes.19.020144}{Phys.\ Rev.\ Phys.\ Educ.\ Res.} \textbf{19}, 020144 (2023). Probes what it means to ``see'' quantum effects from instructor and student perspectives, with implications for aligning learning goals with choice and framing of experiments. (I)

    \item ``Student reasoning about quantum mechanics while working with physical experiments,''  V.\ Borish and H.\ Lewandowski. \href{https://doi.org/10.1103/PhysRevPhysEducRes.20.020135}{Phys.\ Rev.\ Phys.\ Educ.\ Res.} \textbf{20}, 020135 (2024). Features think-aloud interviews with two pairs of students working through single-photon quantum optics experiments. The same experiments elicited different sets of resources from each group of students (wavefunctions and wave phenomena vs.\ information-theoretic interpretations) showing that connections between quantum optics experiments and QISE learning goals may require explicit scaffolding. (I)

    \item ``Affordances and challenges of incorporating a remote, cloud-accessible quantum experiment into undergraduate courses,''  V.\ Borish and H.\ Lewandowski. \href{https://doi.org/10.1103/PhysRevPhysEducRes.21.010133}{Phys.\ Rev.\ Phys.\ Educ.\ Res.} \textbf{21}, 010133 (2025). This study, based on Infleqtion's Oqtant Bose-Einstein Condensate experiment\footnote{Infleqtion has since shut down Oqtant, but the results of this study should be generalizable to similar future products.}, shows that while cloud-based interactive experiments can replicate some of the advantages of in-person labs, challenges with uptime and an inability to replicate the hands-on experience may lead instructors to reconsider their use. (I)

\end{enumerate}

\subsection{Research on attitudes and beliefs}

\add{Students' attitudes and beliefs about QISE are arguably as important as content mastery in determining whether students pursue further research or careers in the quantum industry, motivating a number of studies:}

\begin{enumerate}[resume]

    \item ``Exploratory factor analysis of a precollege quantum information science and technology survey: Exploring career aspiration formation and student interest,'' A.M.\ Kelly \textit{et al.} \href{https://doi.org/10.1140/epjqt/s40507-025-00313-w}{EPJ Quantum Technol.} \textbf{12}, 11 (2025). Investigates how demographics and prior coursework influence precollege students’ interest in QISE and their career aspirations, providing actionable insights for educators and policymakers aiming to support early engagement and broaden participation in quantum education. (E)

    \item \newentry{``Formation of students' interests in quantum technology across STEM majors,'' E.\ Watts and B.M.\ Zwickl. \href{https://doi.org/10.1119/perc.2025.pr.Watts}{Proc.\ 2025 Phys.\ Educ.\ Res.\ Conf.}, 441-446 (2025). An accessible qualitative study featuring interviews with 22 undergraduates, and identifying crosscutting themes among high- vs.\ low-interest students. (E)}

    \item ``Science, technology, engineering, and mathematics undergraduates’ knowledge and interest in quantum careers: Barriers and opportunities to building a diverse quantum workforce,'' J.L.\ Rosenberg, N.\ Holincheck, and M.\ Colandene. \href{https://doi.org/10.1103/PhysRevPhysEducRes.20.010138}{Phys.\ Rev.\ Phys.\ Educ.\ Res.} \textbf{20}, 010138 (2024). Identifies sources of students’ quantum knowledge and barriers to pursuing quantum careers, providing insights useful for educators and researchers aiming to design interventions that promote diversity and engagement in the quantum workforce. (I)

    \item ``Education for expanding the quantum workforce: Students' perceptions of the quantum industry in an upper-division physics capstone course,''  K.A.\ Oliver \textit{et al.} \href{https://doi.org/10.1103/PhysRevPhysEducRes.21.010129}{Phys.\ Rev.\ Phys.\ Educ.\ Res.} \textbf{21}, 010129 (2025). Examines students’ perceptions of the quantum industry through a university-industry capstone project, providing useful insights for educators and instructional designers. (I)

    \item ``Student attitudes toward quantum information science and technology in a high school outreach program,'' M.\ Darienzo \textit{et al.} \href{https://doi.org/10.1103/PhysRevPhysEducRes.20.020126}{Phys.\ Rev.\ Phys.\ Educ.\ Res.} \textbf{20}, 020126 (2024). This work uses a QIST-specific attitudes survey to assess high school students’ affective outcomes from a summer QISE outreach program. It offers valuable guidance for designing effective outreach programs. (A)

    \end{enumerate}

    \add{Instructors' perceptions of QISE are also important, particularly in K-12 where teacher buy-in is critical:}

    \begin{enumerate}[resume]

    \item ``Quantum science and technologies in K-12: Supporting teachers to integrate quantum in STEM classrooms,'' N.\ Holincheck \textit{et al.} \href{https://doi.org/10.3390/educsci14030219}{Educ.\ Sci.} \textbf{14}(3), 219 (2024). Identifies K-12 teachers’ perceived barriers to integrating quantum concepts and technologies into STEM classrooms. It offers valuable insights for designing effective professional development and support programs for educators. (E)

     \item ``Quantum information science and technology professional learning for secondary science, technology, engineering, and mathematics teachers,'' A.M.\ Kelly \textit{et al.} \href{https://doi.org/10.1103/PhysRevPhysEducRes.20.020154}{Phys.\ Rev.\ Phys.\ Educ.\ Res.} \textbf{20}, 020154 (2024). Examines middle and high school STEM teachers’ self-efficacy, QISE knowledge, pedagogical confidence, and awareness of QISE-related careers and pathways. It provides valuable guidance for developing professional learning programs to prepare teachers for integrating quantum concepts into secondary STEM classrooms. (A)

\end{enumerate}

\subsection{Research on K-12 quantum education}

\add{While resources related to K-12 quantum education are interspersed throughout this Resource Letter, the following studies are especially important for understanding the needs and resources of pre-college learners:}

\begin{enumerate}[resume]

    \item ``Preparing precollege students for the Second Quantum Revolution with core concepts in quantum information science,'' C.\ Singh, A.\ Levy, and J.\ Levy. \href{https://doi.org/10.1119/5.0027661}{Phys.\ Teach.} \textbf{60}, 639-641 (2022). Advocates for integrating QIS into K–12 curricula using a compare-and-contrast approach between classical and quantum concepts, offering practical guidance on engaging students and inspiring their interest in quantum careers. (E)

    \item ``Review of literature on quantum information science and technology programs for high school students,'' M.\ Darienzo and A.M.\ Kelly. \href{https://doi.org/10.1109/QCE60285.2024.20464}{Proc.\ 2024 IEEE International Conference on Quantum Computing and Engineering (QCE)} \textbf{3}, 96-103 (2024).\add{A comprehensive guide to the literature on K-12 quantum education as of mid-2024. (E)}

    \item ``Quantum information science and technology high school outreach: Conceptual progression for introducing principles and programming skills,'' D.\ Schneble, T.-C.\ Wei, and A.M.\ Kelly. \href{https://doi.org/10.1119/5.0211535}{Am.\ J.\ Phys.} \textbf{93}, 88-97 (2025). Presents a replicable model for teaching high school students QISE, designed to engage learners without assuming extensive prior knowledge in physics, chemistry, mathematics, or computer science. (E)

    \item \newentry{``Quantum readiness in Latin American high schools: Curriculum compatibility and enabling conditions,'' A.C.\ Alvarado León \textit{et al.} \href{https://arxiv.org/abs/2512.16257}{arXiv:2512.16257} (2025). A detailed analysis of the relative opportunities and constraints around integrating quantum technologies into existing high school curricula across 6 countries. Their methodology provides a useful framework for understanding the enabling conditions and strategies associated with retrofitting quantum into curricula around the world. (E)}

    \item ``Teaching quantum computing to high-school-aged youth: A hands-on approach,'' P.P.\ Angara \textit{et al.} \href{https://doi.org/10.1109/TQE.2021.3127503}{IEEE Trans.\ Quantum Eng.} \textbf{3}, 3100115 (2022). Describes the design and implementation of quantum computing workshops for high school students, integrating unplugged learning activities in both online and in-person formats. (I)
   
    \item ``Making the quantum world accessible to young learners through Quantum Picturalism: An experimental study,'' S.\ Dündar-Coecke \textit{et al.} \href{https://arxiv.org/abs/2504.01013}{arXiv:2504.01013} (2025). Investigates Quantum Picturalism (QPic), a fully diagrammatic approach to qubit quantum mechanics aimed at reducing mathematical barriers for high school learners. It provides actionable insights for educators seeking innovative methods to introduce advanced QISE concepts at early educational stages. (I)
    \begin{itemize}
        \item \newentry{An abbreviated summary of the findings is available at \href{https://arxiv.org/abs/2512.00141}{arXiv:2512.00141} (2025). (E)}
    \end{itemize}

     \item ``Enhancing high school students’ understanding and attitude towards quantum mechanics through discipline-culture framework and cognitive apprenticeship,'' V.V.\ Nautiyal \textit{et al.} \href{https://doi.org/10.1140/epjqt/s40507-025-00407-5}{EPJ Quantum Technol.} \textbf{12}, 104 (2025). Evaluates a fully online Quantum Education STEM \& Research Internship Program (SRIP) for Filipino high school students, offering valuable insights for designing online work-immersion programs. The findings are broadly applicable across international educational contexts. (I)

\end{enumerate}

\subsection{Research on informal QISE education}

\add{Not all QISE education takes place in the classroom. The following studies investigate learning in nontraditional contexts (e.g.~museums, outreach programs):}

\begin{enumerate}[resume]

    \item ``A multidisciplinary, artistic approach to broadening the accessibility of quantum science,'' S.\ Chitransh \textit{et al.} \href{https://doi.org/10.1109/QCE53715.2022.00095}{Proc.\ 2022 IEEE Int.\ Conf.\ Quantum Comput.\ Eng.}, 701-708 (2022). Reports on a novel initiative using theatre and games to make quantum concepts accessible to non-scientific audiences. (E)
    
    \item ``Culturo-scientific storytelling,'' S.\ Goorney \textit{et al.} \href{https://doi.org/10.3390/educsci12070474}{Educ.\ Sci.} \textbf{12}(7), 474 (2022). Introduces a framework for storytelling - developing a narrative for QISE in outreach activities, they call the culturo-scientific storytelling (CSS), arguing that QISE has its own distinct culture and dialogues. (E)
    
    \item ``Educating to the ``culture'' of quantum technologies: A survey study on concepts for public awareness,'' Z.\ C.\ Seskir, S.\ R.\ Goorney, and M.\ L.\ Chiofalo. \href{https://doi.org/10.20897/ejsteme/14193}{Eur.\ J.\ STEM Educ.} \textbf{9}(1), 03 (2024). Investigates which quantum concepts educators prioritize for outreach using a discipline-culture framework, revealing tensions in how quantum technologies relate to physics and advising a reordered emphasis (e.g. qubit over spin) for public engagement. (E)

    \item ``Why teach quantum on your own time: The values of grassroots organisations involved in quantum technologies education and outreach,'' U.\ Genenz \textit{et al.} \href{https://doi.org/10.1140/epjqt/s40507-025-00345-2}{EPJ Quantum Technol.} \textbf{12}, 44 (2025). This article examines the contribution of grassroots organisations to QISE education, characterised by volunteer participation, bottom-up structure and non-profit nature. (E)

    \item ``Quantum at a music festival: The impact of an exhibit about quantum science and technologies on festival visitors,'' V.\ Koeman \textit{et al.} \href{https://arxiv.org/abs/2507.13010}{arXiv:2507.13010} (2025). Evaluate a large-scale public outreach experiment on quantum technologies conducted at the 2024 Lowlands music festival in the Netherlands. Visitors explored Quantum: The Pop-Up Exhibit, a bilingual interactive installation. The authors advocate for evidence-based evaluation of outreach as quantum technologies enter broader public discourse. (E)
\end{enumerate}

\subsection{Research on QISE pedagogy and content}

While diverse in their specific audiences, these studies all attempt to answer variants of the question, ``What content should we be teaching our students and how?''

\begin{enumerate}[resume]

    \item ``From Cbits to Qbits: Teaching computer scientists quantum mechanics,'' N.D.\ Mermin. \href{https://doi.org/10.1119/1.1522741}{Am.\ J.\ Phys.} \textbf{71}, 23-30 (2003). Perhaps the first paper to propose that relatively little traditional quantum mechanics knowledge is needed to learn quantum computing. Included largely for its historical importance; some outdated terminology. (E)

    \item ``The challenge and opportunities of quantum literacy for future education and transdisciplinary problem solving,'' L.\ Nita \textit{et al.} \href{https://doi.org/10.1080/02635143.2021.1920905}{Res.\ Sci.\ Technol.\ Educ.} \textbf{41}(2), 564-580 (2021). An early discussion notable for its introduction of the concept of ``quantum literacy'' as a goal for popular QISE education. (E)

    \item \newentry{``Capturing expert mental models of quantum sensing using concept maps,'' N.\ Pradeep and B.M.\ Zwickl. \href{https://doi.org/10.1119/perc.2025.pr.Pradeep}{Proc.\ 2025 Phys.\ Educ.\ Res.\ Conf.}, 356-261 (2025). This article discusses how researchers come to identify core content in quantum sensing, a domain that lacks traditional teaching materials and is seldom covered in other QISE education research. It is worth watching this team for future research-based pedagogical materials for teaching quantum sensing. (E)}

    \item ``Towards a quantum technology PCK for teachers,'' L.\ Verbraeken \textit{et al.} \href{https://doi.org/10.1088/1742-6596/2750/1/012045}{J.\ Phys.:\ Conf.\ Ser.} \textbf{2750} (GIREP 2022), 012045 (2024). Provides a series of toolkits and concept maps on why and how to teach QISE principles in secondary school education, based on the practical lived experiences of educators in the classroom. (Article: I, resources: E)

    \item ``The core of secondary level quantum education: A multi-stakeholder perspective,'' A.\ Merzel \textit{et al.} \href{https://doi.org/10.1140/epjqt/s40507-024-00237-x}{EPJ Quantum Technol.} \textbf{11}, 27 (2024). Reports on the findings of the first round of a global study of physics researchers, education researchers, and high school teachers on the key concepts and illustrations used in secondary school teaching (both quantum mechanics and QISE), finding a few areas of consensus across profession groups but also marked differences. (I)

    \item ``Introductory quantum information science courses at US institutions: Content coverage,'' J.C.\ Meyer \textit{et al.} \href{https://doi.org/10.1140/epjqt/s40507-024-00226-0}{EPJ Quantum Technol.} \textbf{11}, 16 (2024). A detailed analysis of the content covered in university-level introductory QISE courses broken down by topic, identifying a number of areas of broad consensus alongside a few noteworthy differences by academic discipline and/or level. (I)

\end{enumerate}

Two related studies focused on the important topic of quantum state representations and may be of particular interest to instructors seeking to enliven their curricula and explore the affordances and drawbacks of different representations:

\begin{enumerate}[resume]
        \item ``Investigating the use of multiple representations in university courses on quantum technologies,'' E.\ Rexigel \textit{et al.} \href{https://doi.org/10.1140/epjqt/s40507-025-00327-4}{EPJ Quantum Technol.} \textbf{12}, 22 (2025). Analyzes the symbolic and graphical representations used by 25 instructors to represent qubit states, finding that apart from a few standard symbolic representations there was little consensus particularly for multiple qubits. Some instructors eschewed multi-qubit graphical representations entirely due to perceived lack of options. (I)
        
        \item ``Exploring the mechanisms of qubit representations and introducing a new category system for visual representations: Results from expert ratings,'' L.\ Qerimi \textit{et al.} \href{https://doi.org/10.1140/epjqt/s40507-025-00346-1}{EPJ Quantum Technol.} \textbf{12}, 45 (2025). Investigates the relative affordances and drawbacks of various representations and combinations of representations in terms of factors such as salience, cognitive load, and tendency to address or reinforce common student misconceptions. Of particular relevance to curriculum developers. (A)

\end{enumerate}

\section{Materials for QISE educators}

\subsection{Textbooks}

\add{While consensus on a canonical QISE textbook has yet to emerge among instructors \cite{Meyer:2022}, the following textbooks are recommended as a starting point. Unlike the remainder of this Resource Letter, the resources rated ``A'' here do not require knowledge of education research \textit{per se} but simply denote a higher-level (though accessible) treatment most suitable for a graduate introductory course or second undergraduate course.}

\begin{enumerate}[resume]

    \item \newentry{``Q is for Quantum: Quantum mechanics for those who know only basic arithmetic,'' T.\ Rudolph (2017). An accessible introduction to basic quantum computing with virtually no math prerequisites. Useful particularly for teaching quantum literacy to high school and non-science students who do not require a technically rigorous, quantitative treatment. However, the notation and terminology do not transfer easily to more advanced topics. (E)}

    \item ``Introduction to Classical and Quantum Computing,'' T.\ G.\ Wong. (2021). Introduces a freely accessible textbook bridging classical computation and the mathematical foundations of quantum algorithms, designed for undergraduates and self-learners. (E)

    \item ``Quantum in Pictures: A New Way to Understand the Quantum World,'' B.\ Coecke and S.\ Gogioso. Cambridge Quantum (2023). Offers a visually intuitive, low-mathematics introduction to quantum theory by combining imagery and conceptual narration to broaden accessibility for newcomers. (E)

    \item ``Quantum Computer Science: An Introduction,'' N.\ David Mermin. Cambridge University Press (2007). Provides a clear, minimal-mathematics introduction to quantum computation for computer scientists and mathematicians, treating key algorithms, error correction, and foundations without delving into hardware or deep physics. (E-I)

    \item ``Quantum Computation and Quantum Information,'' M.\ A.\ Nielsen and I.\ L.\ Chuang. Cambridge University Press (10th Anniversary Edition, 2010). A foundational and comprehensive reference that combines rigorous treatment of quantum algorithms, error correction, and information theory with pedagogy and exercises, widely used by advanced students and researchers. (I\add{-A})

    \item ``Quantum Programming in Depth: Solving Problems with Q\# and Qiskit,'' M.\ Mykhailova. \href{https://www.manning.com/books/quantum-programming-in-depth}{Manning, 2025}. Advances practical quantum software development by guiding readers through algorithm implementation, testing, and performance evaluation with Q\# and Qiskit. (A)

\end{enumerate}

\subsection{Research-based curricular materials}

Research-based curricular materials are based on research into student reasoning and difficulties and validated through a rigorous process to ensure their effectiveness. Most use various modalities of active learning pedagogies (e.g.\ clicker questions, tutorials, simulations). They are ideal for instructors who wish to enjoy the benefits of research-based pedagogies without extensive preparation.

\begin{enumerate}[resume]
    \item \textbf{Quantum Interactive Learning Tutorials (QuILTs)}, C.\ Singh \textit{et al.}: \url{https://www.physport.org/curricula/QuILTs/}. Adaptable, interactive, research-based tutorials intended for use in breakout groups during active-learning courses or discussion sections. Includes worksheets, facilitation instructions and in some cases, simulation resources. Developed in a physics context so may require tweaks to notation. Includes both traditional quantum mechanics and QISE topics. (E)
    
    QISE-related QuILTs have been featured in a number of published research articles, including:
    
    \begin{itemize}
        \item ``Interactive learning tutorial on quantum key distribution,'' S.\ DeVore and C.\ Singh. \href{https://doi.org/10.1103/PhysRevPhysEducRes.16.010126}{Phys.\ Rev.\ Phys.\ Educ.\ Res.} \textbf{16}, 010126 (2020). Demonstrates the effectiveness of a tutorial on the B92 quantum key distribution protocol including detailed pre-post test data. (I)

        \item ``Investigating and improving student understanding of the basics of quantum computing,'' P.\ Hu, Y.\ Li, and C.\ Singh. \href{https://doi.org/10.1103/PhysRevPhysEducRes.20.020108}{Phys.\ Rev.\ Phys.\ Educ.\ Res.} \textbf{20}, 020108 (2024). 
        A tutorial on fundamental quantum computing concepts was designed around documented student difficulties and iteratively refined through research. Findings showed notable learning gains beyond traditional lecture-based instruction, with only a few persistent challenges. (I)

        \item ``Student understanding of the Bloch sphere,'' P.\ Hu \textit{et al.} \href{https://doi.org/10.1088/1361-6404/ad2393}{Eur.\ J.\ Phys.} \textbf{45}(2), 025705 (2024). This study investigates undergraduate students’ understanding of the Bloch sphere following the implementation of a research-based tutorial. Findings indicate notable improvements in students’ conceptual clarity and performance, particularly in interpreting measurement probabilities and identifying equivalent states. (I)
    \end{itemize}
    
    \item \textbf{Quantum Mechanics Visualization Project (QuVIS)}, A.\ Kohnle \textit{et al.}: \url{https://www.st-andrews.ac.uk/physics/quvis/}. Research-validated simulation resources on both traditional quantum mechanics and QISE topics, many with associated editable tutorials, in both English and German. Read ``For instructors'' before use. (E)
    
    QuVIS has been validated in a number of relevant research-based publications most notably:
    \begin{itemize}
        \item ``Enhancing student learning of two-level quantum systems with interactive simulation,'' A.\ Kohnle \textit{et al.} \href{https://doi.org/10.1119/1.4913786}{Am.\ J.\ Phys.} \textbf{83}(6), 560-566\add{ (2015)}. (E) 
    \end{itemize}
    
    \item \textbf{AceQIS: Tutorials About Quantum Information Science}, G.\ Corsiglia \textit{et al.}: \url{https://acephysics.net/qis}. A new and rapidly-growing set of web-based tutorials that provide interactive feedback based on student answers mimicking a TA. Can be used similar to small-group paper tutorials in class (ideal) or assigned as homework problems in courses where time constraints and/or classroom layout would preclude tutorial use.\footnote{QISE-specific tutorials have not yet been featured in peer-reviewed publications, but Corsiglia's thesis \cite{Corsiglia:thesis} discusses their design philosophy in depth and includes evidence of their effectiveness both in-person (highest) and as homework problems (still substantial).} (E)

    \item \textbf{PhET}, a globally-recognized simulations project whose work has been translated into dozens of languages, has started to also release simulations related to QISE, with the most relevant at time of writing being a tutorial on quantum measurement: \url{https://phet.colorado.edu/en/simulations/quantum-measurement} (E).
    
    Simulations have been validated in a number of publications, including (not QISE-specific):
    \begin{itemize}
        \item ``Developing and researching PhET simulations for teaching quantum mechanics,'' S.\ McKagan \textit{et al.} \href{https://doi.org/10.1119/1.2885199}{Am.\ J.\ Phys.} \textbf{76}(4), 406-417 (2008). (E)
    \end{itemize}

    \item ``Designing and implementing materials on quantum computing for secondary school students: The case of teleportation,'' S.\ Satanassi, E.\ Ercolessi, and O.\ Levrini. \href{https://doi.org/10.1103/PhysRevPhysEducRes.18.010122}{Phys.\ Rev.\ Phys.\ Educ.\ Res.} \textbf{18}, 010122 (2022). This article, one of several by the same authors, introduces the design philosophy behind a series of research-based curricular materials developed for Italian high-school students and demonstrates their effectiveness through mixed-methods analysis. (I)

    \item ``Toward personalizing quantum computing education: An evolutionary LLM-powered approach,'' I.\ Elhaimeur and N.\ Chrisochoides \del{arXiv:2504.18603}\href{https://doi.org/10.1109/QCE65121.2025.20549}{\add{Proc.~ 2025 IEEE Int.~Conf.~Quantum Comput.~Eng.}} \add{\textbf{3}, 174-184} (2025). An approach for the use of AI models to develop personalized quantum computing lesson plans for students, leveraging Microsoft's Quantum Katas and Copilot LLM tool, combining the expertise of QISE educators and computer scientists. While intriguing in principle, at time of writing few conclusive results are available. (I)
    
\end{enumerate}

\subsection{Assessing student learning in QISE}
\label{sec:assessment}

Research-based assessments enable instructors to reliably benchmark and compare student learning across courses and institutions. They undergo a rigorous evaluation process for validity and reliability, and are based in research on student reasoning and common naive conceptions. Instructors may find these instruments useful for diagnostic purposes, especially when teaching multiple versions of a course over time. They are not intended to be graded on correctness. For more information about research-based conceptual assessments, refer to Resource Letter RBAI-1 \cite{Madsen:2017}.

\begin{enumerate}[resume]
    \item \textbf{Quantum Computing Conceptual Survey (QCCS):}  Validated assessment of quantum computing/information foundations intended for first courses in QISE at the undergrad or graduate level. Administer as a post-test or a mid-course diagnostic once all fundamental topics are covered. (E)
    
    QCCS validation featured in: ``New paradigms in quantum education for the Second Quantum Revolution,'' J.C. Meyer. Ph.D.\ Thesis, \href{https://www.proquest.com/docview/3205639873/abstract/6F5BD6E93A974F00PQ/1}{University of Colorado Boulder} (2025). (A)\add{ A formal validation paper is anticipated in Q2 2026.}

    \item \textbf{Quantum Concept Inventory (QCI):} A similar instrument to QCCS but intended for the K-12 level and in a somewhat earlier phase of development. Keep eyes pealed for future developments. (E)
    
    QCI discussed in: ``Contributions from pilot projects in quantum technology education as support action to Quantum Flagship,'' S.\ Faletič \textit{et al.}. In \href{https://doi.org/10.1007/978-3-031-72541-8_15}{\textbf{Teaching and Learning Physics Effectively in Challenging Times}}, S.\ Faletič and J. Pavlin, eds., 219-238, (Cham: Springer, 2024) (I).

    \item \textbf{Quantum Information Science Concept Inventory Test (QISCIT):} A newly-announced test that has not been tested on students and therefore does not (yet) meet the standard definition of a research-based assessment, but worth watching.

    Discussed in: ``QISCIT: A validated concept inventory assessment for quantum information science,'' K.\ Durkin \textit{et al.}, \href{https://arxiv.org/abs/2506.17122}{arXiv:2506.17122} (2025). (I)
    
\end{enumerate}

\subsection{Quantum games}

\add{Gamification is a useful strategy for quantum awareness and building technical knowledge. The following resources provide a useful starting point for instructors wishing to use such games:}

\begin{enumerate}[resume]
    \item ``Quantum games and interactive tools for quantum technologies outreach and education,'' Z.C.\ Seskir \textit{et al.} \href{https://doi.org/10.1117/1.OE.61.8.081809}{Opt.\ Eng.} \textbf{61}(8), 081809 (2022).  Summarises the state of the art in interactive tools and resources for quantum outreach, including games, simulations, and explorable explanations. This article highlights the great diversity in tools available. (E)

    \item ``Games for quantum physics education,'' M.L.\ Chiofalo \textit{et al.} \href{https://doi.org/10.1088/1742-6596/2727/1/012010}{J.\ Phys.:\ Conf.\ Ser.} \textbf{2727} (3rd World Conf.\ Phys.\ Educ.\ 2021), 012010 (2024). This article discusses the significance of games for quantum education, in particular how they help to address the challenge of creative abstraction present in conceptualising quantum phenomena. (E)

    \item ``Defining quantum games,'' L.\ Piispanen \textit{et al.} \href{https://doi.org/10.1140/epjqt/s40507-025-00308-7}{EPJ Quantum Technol.} \textbf{12}, 7 (2025). Piispanen et al. (2025) take an analytical approach to defining quantum games, by surveying games collected over several years. They identify several common dimensions and propose a definition for quantum games. (E)
    
\end{enumerate}

\subsection{Teaching about ethical and societal implications of quantum technologies}

The advent of the Second Quantum Revolution and the industrialization of quantum technologies comes with the necessity to not just teach students about the technical aspects of quantum technologies but also to think critically about their societal implications -- an emerging field of scholarship known as Quantum Ethics or Responsible Quantum Technologies. The following resources may be especially useful for educators and program developers looking at integrating these topics in their curricula:

\begin{enumerate}[resume]

    \item ``Ethics education in the quantum information science classroom: Exploring attitudes, barriers, and opportunities,'' J.C.\ Meyer, N.\ Finkelstein, and B.R.\ Wilcox.  \href{https://doi.org/10.18260/1-2--40456}{Proc.\ 2022 ASEE Ann.\ Conf.\ Expo.}, 36517 (2022). Reports findings from QISE instructor interviews showing interest in incorporating quantum ethics discussions into the classroom but also perceived barriers to entry. Discusses how these barriers can be reduced. (E)

    \item ``A holistic approach to quantum ethics education,'' J.É.\ Arrow, S.E.\ Marsh, and J.C.\ Meyer. \href{https://doi.org/10.1109/QCE57702.2023.20332}{Proc.\ 2023 IEEE Int.\ Conf.\ Quantum Comput.\ Eng.} \textbf{3}, 119-128 (2023). A detailed primer on quantum ethics education, covering specific ethical issues pertaining to quantum technologies and techniques from ethics education research literature on effective instruction. Also introduces a set of research-inspired modular curricular materials. (E)

    \item ``Talking about responsible quantum: `Awareness is the absolute minimum ... that we need to do,''' T.\ Roberson. \href{https://doi.org/10.1007/s11569-023-00437-2}{NanoEthics} \textbf{17}, 2 (2023). A stakeholder interview study on the importance of awareness and discussion in ensuring ethical and responsible development of quantum technologies. (E)

    \item ``The role of community building and education as key pillar of institutionalizing responsible quantum,'' S.\ Vishwakarma \textit{et al.} \href{https://doi.org/10.1109/QCE60285.2024.10258}{Proc.\ 2024 IEEE Int.\ Conf.\ Quantum Comput.\ Eng.} \textbf{3}, 86-91 (2024). A position paper from IBM on the role of educational initiatives in fostering equitable access to and responsible development of QISE technologies.\footnote{Note that while the educational design principles discussed in the paper are sound, several of IBM's initiatives (e.g.\ the IBM-HBCU Quantum Center) have faced critique by the community for falling short of their advertised promises. It is not our intent to enter this debate.} (E)

    \item ``Building student understanding of quantum information science and engineering through projects on applications to medical technologies,'' J.L.\ Rosenberg and N.\ Holincheck. \del{arXiv:2508.03850}\href{https://doi.org/10.1109/QCE65121.2025.20527}{\add{Proc.~2025 IEEE Int.~Conf.~Quantum Comput.~Eng.}}\add{ \textbf{3}, 82-85} (2025). Discusses opportunities for engagement of students with interest in health and biomedical sciences in QISE through projects focused on applications of quantum technologies in medicine, including ethical and societal implications. (E)

    \item ``Second quantum revolution: The progressive design of an approach to value its cultural and conceptual scope,'' S.\ Satanassi and O.\ Levrini. \href{https://doi.org/10.1103/PhysRevPhysEducRes.21.010112}{Phys.\ Rev.\ Phys.\ Educ.\ Res.} \textbf{21}, 010112 (2025). Details the development and refinement of a research-based curriculum blending fundamental principles of quantum technologies with societal and philosophical implications through future- and civic-oriented activities. (I)

\end{enumerate}

\subsection{Teaching about quantum amid media hype}

\add{A related concern for many educators is the phenomenon of media hype. The following articles will be useful for educators wishing to better understand the quantum ``hype cycle'' and how to respond to its consequences in their teaching.}

\begin{enumerate}[resume]

    \item ``Is everything quantum ‘spooky and weird’? An exploration of popular communication about quantum science and technology in TEDx talks,'' A.L.\ Meinsma \textit{et al.} \href{https://doi.org/10.1088/2058-9565/acc968}{Quantum Sci.\ Technol.} \textbf{8}, 035004 (2023). The authors conduct a content analysis of 501 TEDx talks to show that “spooky” framing appears in about a quarter of quantum-themed presentations, experts more often explain quantum concepts than nonexperts, and benefit-centric narratives dominate over risk framing. (E)

    \item ``How media hype affects our physics teaching: A case study on quantum computing,'' J.C.\ Meyer \textit{et al}. \href{https://doi.org/10.1119/5.0117671}{Phys.\ Teach.} \textbf{61}(5), 339-342 (2023). Examines how sensationalized media portrayals of quantum science influence student expectations and misconceptions, and proposes instructional strategies to counteract those distortions. (E)

    \item ``Quantum Computing in the NISQ era and beyond,'' J.\ Preskill. \href{https://doi.org/10.22331/q-2018-08-06-79}{Quantum} \textbf{2}, 79 (2018). Provides a foundational perspective on the NISQ era, arguing that near-term quantum devices with 50–100 qubits will enable explorations of many-body physics despite noise constraints, and emphasizes the necessity of advancing toward fault-tolerant quantum computing. \add{We recommend this article for anyone wishing to separate the facts from the hype.} (I)
    
    \item ``Strategies educators can use to counter misinformation related to the quantum information revolution,'' J.S.\ Kashyap and C.\ Singh. \href{https://doi.org/10.1088/1361-6552/adbeb1}{Phys.\ Educ.} \textbf{60}, 035024 (2025). Proposes concrete instructional strategies for confronting misinformation and misconceptions about quantum information science, grounded in research-based pedagogy to promote accurate understanding. (I)

    \item ``Like a coin spinning in the air: the effect of (non-)metaphorical explanations on comprehension and attitudes towards quantum technology,'' A.L.\ Meinsma, W.\ G.\ Reijnierse, and J.\ Cramer. \del{arXiv:2506.10539 (2025)}\href{https://doi.org/10.1075/msw.25018.mei}{Metaphor Soc.\ World} \add{(2026)}. Experimentally compares metaphorical, literal, and no explanations of a quantum phenomenon, showing both metaphorical and literal explanations increase comprehension but reduce perceived comprehension, with negligible net effect on attitudes—implying metaphors confer limited communicative benefit. (I)

\end{enumerate}

\subsection{Miscellaneous resources}

\add{Other materials this Resource Letter would be incomplete without:}

\begin{enumerate}[resume]

    \item Quantum Odyssey (educational game), Quarks Interactive. \href{https://www.quarksinteractive.com/quantum-odyssey}{quarksinteractive.com/quantum-odyssey}. L.\ Nita, N.\ Chancellor, L.\ Mazzoli Smith, H.\ Cramman, and G.\ Dost. \href{https://arxiv.org/abs/2106.07077}{arXiv:2106.07077} (2021) An educational puzzle game that conveys quantum computing logic through visual, interactive challenges rather than mathematics or code, making foundational quantum concepts accessible to general audiences and students. (E)

    \item Qubit by Qubit. \href{https://www.qubitbyqubit.org/}{qubitbyqubit.org}. Offers K-16 and workforce quantum education programs, including year-long high school courses, internships, workshops, and policy advocacy, with a strong focus on accessibility, equity, and real-world training; by 2023 it had engaged over 22,500 learners across all U.S. states. (E-I)

    \item QWorld (Association) \href{https://qworld.net}{qworld.net}. A global non-profit network that coordinates open quantum tutorials, workshops, internships, and local groups with an emphasis on expanding access, grassroots participation, and capacity building. Their resources are freely available on: \href{https://gitlab.com/qworld}{gitlab.com/qworld} (E-I-A)

    \item ``Quantum computing with Qiskit,'' A.\ Javadi-Abhari \textit{et al.} \href{https://arxiv.org/abs/2405.08810}{arXiv:2405.08810} (2024). Provides a deep overview of the Qiskit software development kit—covering its design decisions, architecture, core components, and ecosystem—and demonstrates a complete workflow highlighting its flexibility, scalability, and retargetability for quantum applications. (I)

    \item The Virtual Quantum Optics Laboratory (\href{https://www.vqol.org/}{VQOL}) B.\ R.\ La Cour \textit{et al.} \href{https://doi.org/10.1109/QCE53715.2022.00091}{Proc.\ 2022 IEEE Int.\ Conf.\ Quantum Computing and Engineering}, 677–687 (2022). VQOL is a web-based quantum optics simulation environment enabling design and execution of realistic experiments with optical components and measurement models, and shows it can faithfully replicate many phenomena, making it valuable for both instruction and research. (I-A)

\end{enumerate}

\section{Conclusions}

While the 2025 International Year of Quantum Science and Technology \del{is coming}\add{has come} to a close, it is our hope that research on QISE education is just beginning. Much remains to be learned in this field, both through traditional education research methodologies and the in-the-field experiences of educators. It is our hope that this Resource Letter will not only help make this literature accessible to practitioners but serve as a valuable entry point for new researchers to build on.

\subsection{Outlook for future research}

In many ways, the future of QISE education research appears very bright. 
A few particularly exciting developments lie on the horizon:

\begin{itemize}
    \item \textbf{Research-based assessment outcomes:} Research-based assessment instruments (Sec.~\ref{sec:assessment}) will be a game-changer for quantum education research, both by providing a benchmark for validating novel curricular materials and for opportunities to mine the rich quantitative datasets provided.
    \item \textbf{AI for quantum education research:} The AI revolution in education driven by large-language models such as ChatGPT coincides with the rapid deployment of QIS coursework around the world. QISE education programs may prove an ideal testbed for integrating these technologies into the classroom given the flexibility of emerging curricula and the challenges of interdisciplinary education particularly at the postsecondary level. AI may also enable advances in qualitative and quantitative data analysis in education research.
\end{itemize}

\subsection{Risks and a call to action}

While new and exciting developments in QISE education research are undoubtedly on the way, we also want to use this opportunity to discuss looming storm clouds on the horizon for quantum education research. As a field, it is important that we seize the current moment when programs are still in their formative stages so that research-based best practices are identified and promulgated from the beginning. It is much easier to develop QISE education programs intentionally from the beginning than to retrofit pedagogical best practices into entrenched classroom and disciplinary environments. To ensure the positive developments discussed in this article continue, we urge action to address the following barriers to realizing the benefits of QISE education and quantum education research:

\begin{itemize}
    \item \textbf{Disciplinary insularity}: While QISE is an interdisciplinary field, much of the education research still remains in the confines of individual disciplines. To maximize scholarly productivity and avoid redundant work, more effort is necessary to bring together scholars from different disciplinary backgrounds. This collaboration must also include scholars from technical disciplines and social sciences.

    \item \textbf{Research equity}: As discussed in Sec.~\ref{sec:missing}, most of the research studies listed in this Review Letter were conducted on relatively elite student populations in the Global North, a trend warranting methodological consideration in future studies \cite{Kanim:2020}.

    \item \textbf{Journal availability}: With the exception of the annual Quantum Science and Engineering Education Conference (QSEEC) proceedings\footnote{Published as a separate volume within the proceedings of the IEEE International Conference on Quantum Computing and Engineering, where QSEEC is co-located.}, there remain few dedicated interdisciplinary spaces for the publication of quantum education research. Most articles we identified were published within traditional physics education outlets (titles such as \textit{Physical Review Physics Education Research}, \textit{American Journal of Physics}, \textit{European Journal of Physics}, \textit{Physics Education}, \textit{The Physics Teacher}, and GIREP/PERC proceedings), with much fewer in engineering education journals such as \textit{IEEE Transactions on Education} and the American Society of Engineering Education (ASEE) conference proceedings. 

    A particular risk is the impending conclusion of the \textit{EPJ Quantum Technology} special section on Quantum Education. At time of writing, this collection has published 29 journal articles from researchers across education disciplines and is one of the community's most well-known publication platforms.\footnote{Special sections in \textit{Optical Engineering} (2022) and \textit{Quantum Science and Technology} (2021-2024) have also been important venues for QISE education research publication.} The success of this special section illustrates the importance of a dedicated QISE education journal. 

    \item \textbf{Geopolitical considerations:} The rise in anti-intellectualist political movements across the globe presents a longer-term risk for quantum education research. The US, once a global epicenter for STEM education research, is proposing and implementing drastic cuts to STEM education research funding. As US leadership falters, we urge other regions to step up to  continue funding this important work.

\end{itemize}

\section{Acknowledgment}

We thank the American Journal of Physics for providing seed funding and impetus for this project. The authors have no conflicts to disclose, apart from inevitable self-interests in the choices of resource to include (mitigated to the extent possible through the article selection process described in Sec.~\ref{sec:inclusion}).

\bibliography{AJP_Resource_Letter_QIE}

\end{document}